\documentclass[aps,pra,reprint,groupedaddress]{revtex4-1}

\usepackage{amssymb,amsmath}
\usepackage[bookmarks = true, pdfpagemode = None, pdfstartview = FitH, colorlinks = true, urlcolor = blue]{hyperref}
\usepackage{graphicx}

\input{Qcircuit.tex}
\renewcommand{\Qcircuit}[1][0em]{\xymatrix @*=<#1>}
\newcommand{\cmeasure}[1]{*+[F-:<.9em>]{#1} \cw}

\newcommand{\kb}[1]{\ket{#1}\bra{#1}}

\begin{document}


\title{Optimal eavesdropping on QKD without quantum memory}


\author{Aur\'elien~Bocquet}
\email[]{aurelien.bocquet@telecom-paristech.fr}
\affiliation{Institut Telecom / Telecom ParisTech \& CNRS LTCI, 75013 Paris, France}


\author{Anthony~Leverrier}
\affiliation{ICFO-Institut de Cienc\`es Fot\`oniques, 08860 Castelldefels (Barcelona), Spain}
\author{Romain~All\'eaume}
\affiliation{Institut Telecom / Telecom ParisTech \& CNRS LTCI, 75013 Paris, France}

\date{July 23rd, 2011}

\begin{abstract}
We consider the security of the BB84, six-state and SARG04 quantum key distribution protocols when the eavesdropper doesn't have access to a quantum memory. In this case, Eve's most general strategy is to measure her ancilla with an appropriate POVM designed to take advantage of the post-measurement information that will be released during the sifting phase of the protocol. After an optimization on all the parameters accessible to Eve, our method provides us with new bounds for the security of six-state and SARG04 against a memoryless adversary. In particular, for the six-state protocol we show that the maximum QBER for which a secure key can be extracted is increased from $12.6\%$ (for collective attacks) to $20.4\%$ with the memoryless assumption.
\end{abstract}


\maketitle

\section{Introduction}
Following the invention of quantum key distribution (QKD) and of its first protocol \cite{bb84}, a central issue in QKD theory has been to find sets of assumptions under which formal security proofs could be derived. In this perspective, since Alice and Bob act as honest players, only the unpredictable behavior of the attacker Eve remains to be captured. Defining a security model thus essentially reduces to making simplifying assumptions allowing to bound the attacking capabilities of Eve. However, for a security model to be of interest, it also needs to fulfill several additional constraints and in particular to allow a tractable derivation of security proofs while presenting a level of generality ideally as large as possible.\\

Intercept-resend (IR) attacks \cite{RMP_gisin}, are arguably the simplest and the first attacks that have been considered \cite{bb84}. In this security model, Eve has in particular no quantum memory and her strategy consists in making an immediate measurement on a fraction of the individual quantum states sent by Alice, and then to resend to Bob, for each individual measurement, the quantum state corresponding to the eigenstate of her measurement result. 
IR attacks can be optimized \cite{bechmann} and have the notable interest of being implementable with current technology \cite{lodewick,curty}  since Eve is essentially playing a role similar to Bob.\\

IR attacks are however not very general and proving the security of QKD within stronger security models has rapidly attracted most of the attention of researchers.
This has been especially true concerning the search for an unconditional security proof of QKD, i.e. a proof valid against the most general quantum attacker. The important theoretical efforts that have been invested in this direction however proved that this was not easy, and if BB84 \cite{mayers_1998,renner_phd,shor-2000-85}  and several other QKD protocols  \cite{lo_01,cirac,tamaki,ekert91} have been proven secure against  the most general quantum attacker, it is however not the case yet for most protocols. For this reason, weaker security models, that dates back from the initial categorization of security proofs \cite{RMP_gisin}, namely individual attacks and collective attacks still play key roles as security models in QKD and that have an important feature in common: they rely on the assumption that Eve is in possession of a quantum memory.\\

The assumption about the availability of a quantum memory can however be challenged in practice. Recent results on implementations of quantum memory \cite{qmem,RMP_mem} confirm that it is still technologically very hard to design and build a reliable one: more precisely a high fidelity quantum memory with an arbitrary long storage time doesn't exist yet. In the case of QKD, it is therefore reasonable in a realistic setting to consider the adversary to be memoryless: indeed, the honest participants don't need a quantum memory to perform a QKD protocol, so that they can always wait long enough for the eavesdropper memory to be completely noisy and useless. Studying the security of a QKD protocol in the memoryless adversary model is moreover useful to quantitatively assess the influence of the ``memoryless assumption''. As explained above, this assumption is realistic from a technological point of view but nevertheless leads to weakening the security model with respect to individual attacks and of course stronger attacks. The explicit derivation of the secure key rate under the optimal memoryless attack allows to evaluate what can be seen as a ``memoryless trade-off'', namely the performance gain versus the weakening of the security model. \\

Despite this practical interest, academic works on the subject are scarce. Since most of QKD security proofs have so far been conducted under one of the main security models (individual, collective and coherent attacks), it was always assumed that Eve had a quantum memory and memoryless attacks on QKD have not been studied widely. For the case of the six-state \cite{bruss_98} and SARG04 protocols \cite{SARG04}, the optimal memoryless attacks have not been studied to our knowledge. For the BB84 protocol, one of the analyses on the subject \cite{lutkenhaus_97} studied the optimal POVM that Eve could use to measure the qubit flying from Alice to Bob and found that a key could no longer be extracted for a QBER greater than $15.4\%$ against a memoryless adversary. In our work, we confirm the optimality of this previously known bound on BB84 and provide new tight bounds for the security of the six-state and SARG04 protocols against a memoryless adversary.

After a short description of the QKD protocol considered, we discuss the construction of the attack model. We then compute the secret key rate and optimize it over all parameters before applying the method to the BB84, six-state and SARG04 QKD protocols.

\section{Description of the protocol}

To avoid the use of unnecessarily complicated notations, we describe our attack by using the BB84 protocol with forward reconciliation \cite{RMP_scarani}. The generalization of the attack to the six-state and SARG04 protocols is then straightforward as explained in section \ref{protocols}. Alice and Bob have access to a quantum channel and a classical authenticated channel. The protocol can be decomposed in 4 steps:
\begin{enumerate}
  \item \textbf{Preparation}: For $n\in \mathbb{N}$, Alice chooses randomly $x^n=(x_1,..,x_n)\in\{0,1\}^n$ which represents the raw key, $\theta^n=(\theta_1,..,\theta_n)\in\{0,1\}^n$ which represents the basis, and she prepares the state $\ket{\phi^n}=H^{\theta^n}\ket{x^n}$ before sending it to Bob who measures the state in a random basis. The output of his measurement is $y^n=(y_1,..,y_n)\in\{0,1\}^n$.
  \item \textbf{Sifting}: Alice and Bob publicly announce their choice of basis and discard the instances where the basis disagree. For simplicity (but without loss of generality) we forget the bits when the bases disagree. The resulting sifted raw keys are $x^n$ and $y^n$. Alice and Bob then use a small amount of raw key to estimate the QBER: if it is below a certain value they decide to resume the protocol, or else they abort it.
  \item \textbf{Error correction}: Based on the value of the QBER, Alice computes an error correction message $I_{ec}$ and sends it to Bob. Bob recovers $x^n$ based on $y^n$ and on the information provided by Alice $I_{ec}$.
  \item \textbf{Privacy amplification}: Alice and Bob use a two-universal hashing function to transform their information $x^n$ into a key of size $l$.
\end{enumerate}

The preparation step in the protocol is described above corresponds to a {\it Prepare-and-Measure} (P\&M) scheme: Alice uses a random number generator to prepare a quantum state. This scheme doesn't require Alice and Bob to have a quantum memory and can be easily implemented with today's technology \cite{RMP_scarani}. It is possible to transform this protocol in an {\it entangled-based} (EB) scheme whose security is easier to prove but which requires Alice and Bob to have a quantum memory:  instead of randomly choosing the bits $x$ and $b$ to encode the information in $\ket{\phi}$, Alice can prepare an entangled state $\ket{\Phi}_{AB}=\frac{\ket{0}_A\otimes\ket{0}_B+\ket{1}_B\otimes\ket{1}_B}{\sqrt{2}}$, send half of the state to Bob and finally measure her half in the basis $b$ to get the key string $x$: after this operation is repeated $n$ times,  Bob holds the state $\ket{\phi}$ in his laboratory. Clearly this transformation makes the protocol much harder to implement but the security of the EB scheme implies the security of the easier to implement P\&M scheme \cite{renner_gis}. 

\section{Description of the attack model}
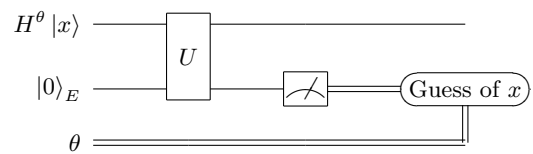
\begin{figure}[htbp]
\begin{center}
\[ \Qcircuit @C=3em @R=1.5em {
    \lstick{H^\theta\ket{x}} & \multigate{1}{U} 	& \qw 	& \qw 	\\
    \lstick{\ket{0}_E} & \ghost{U}                        		& \meter  	& \cmeasure{\mbox{Guess of $x$}}   \\
    \lstick{\theta} & \cw 	& \cw 	&\cw \cwx
}\]
\caption{Quantum circuit representing Eve's attack in the {\sl P\&M} scheme of the BB84 protocol}
\label{circuit}
\end{center}
\end{figure}

We consider the most general actions Eve can perform to gain information with the restriction that she doesn't have a quantum memory. During the preparation phase, Eve is allowed to let an ancilla interact with the qubit flying from Alice to Bob and to measure this ancilla immediately after the interaction with an arbitrary POVM. If $U$ is the unitary interaction applied by Eve to the system, the attack can be represented by the quantum circuit presented in FIG. \ref{circuit}. After the interaction, Eve's ancilla is entangled to the flying qubit and the statistics of any measurement performed on the ancilla can be correlated to the raw key shared by Alice and Bob. Based on the classical information obtained from the measurement and the basis information $\theta$ received afterwards, Eve computes a guess on the raw key bit shared by Alice and Bob. The computation of the probability that her guess is correct is done in the next part of the article.

The attack represented in the P\&M scheme on FIG. \ref{circuit} can also be described in the equivalent EB scheme. In this case, Alice initially prepares a pure state $\ket{\Phi}_{AB}$ and sends one half to Bob through the quantum channel. After the interaction with Eve (or equivalently,  the action of the quantum channel), Alice and Bob share a mixed state $\rho_{AB}=\mathcal{E}(\ket{\Phi}\bra{\Phi}_{AB})=\mbox{Tr}_{E}\ket{\Psi}\bra{\Psi}_{ABE}$ where $\ket{\Psi}_{ABE}$ is a purification of this state. Eve can perform a measurement on her part of the purification $\rho_E=\mbox{Tr}_{AB}\ket{\Psi}\bra{\Psi}_{ABE}$ to gain some information on the secret bit shared by Alice and Bob. We write $X$, $Y$ and $K$ the random variables representing the results of the measurements performed by Alice, Bob and Eve on their part of the purification $\ket{\Psi}_{ABE}$.

\section{Computation of the secret key rate}

In our model, Eve doesn't have a quantum memory and must therefore measure immediately after the interaction the ancilla used in the attack. This means that at the beginning of the classical post-processing of the protocol, Alice, Bob and Eve share a classical probability distribution. The size $l$ of the secret key that can be extracted after the privacy amplification is given by the Csisz\'ar and K\"orner bound \cite{korner}
\begin{eqnarray}
l&=&I_{AB}-\max_{strategies} I_{AE}\\
&=&n[I(X:Y)-\max_{strategies} I(X:K\Theta)],\notag
\end{eqnarray}
where $I_{AB}$ is the mutual informations between Alice and Bob and $\max I_{AE}$ is the maximization on all the eavesdropping strategies of the mutual information between Alice and Eve.

When Eve interacts with the flying qubit she alters it in a way that generates some errors in Bob's string. This perturbation is described by the quantum bit error rate $Q$ measured by Alice and Bob. For a given QBER $Q$, the mutual information between Alice's and Bob's bit string is given by the capacity of the binary symmetric channel
\begin{equation}
I(X:Y)=H(X)-H(X|Y)=1-h(Q),
\end{equation}
where $h(p)=-p\log_2{p}-(1-p)\log_2{(1-p)}$ is the binary entropy. \\

We now compute $I_{AE}$, the mutual information between Alice and Eve. After the reconciliation phase, Eve has access to the result of her measurement $K$ and to the basis information $\Theta$ to guess the value of Alice's bit. We can then write that
\begin{eqnarray}
I(X:K\Theta) & = & H(X)+H(K\Theta)-H(XK\Theta)\\
&=&1+H(K|\Theta)+H(\Theta)-H(K|X\Theta)-H(X\Theta)\notag\\
&=& H(K|\Theta)-H(K|X\Theta)\notag
\end{eqnarray}
where we used the fact that $X$ and $\Theta$ are independent so that $H(X|\Theta)=H(X)=1$. We can then compute the conditional entropies:
\begin{eqnarray}
H(K|X\Theta)&=&\sum_{x,\theta}p(X=x,\Theta=\theta).H(K|X=x,\Theta=\theta)\notag\\
&=&\frac{1}{4}\sum_{k,x,\theta}\Lambda [p(K=k|X=x,\Theta=\theta)]\\
\mbox{and \ }H(K|\Theta) &=& \frac{1}{2}\sum_{k,\theta}\Lambda [p(K=k|\Theta=\theta)]
\end{eqnarray}
where $\Lambda(x)=-x\log_2(x)$. The conditional probabilities used in these formulas will be computed in the next section.\\

Finally we compute the secret key rate $r=l/n$ that can be extracted from the protocol:
\begin{equation}
r = 1-h(Q) -\max_{strategies} [H(K|\Theta)-H(K|X\Theta)].
\end{equation}

\section{Optimization of Eve's attack}
For a given QBER that Eve allows herself to create on the channel between Alice and Bob, we want to maximize $I_{AE}$ over all possible interactions $\mathcal{E}$ and all possible POVMs  with the restriction that she doesn't use a quantum memory. To optimize $I_{AE}$, we need in all generality to take into account two elements:
\begin{itemize}
\item For each target QBER, we need to consider all the possible interactions $U$ that Eve can do. Equivalently in the EB scheme we need to consider all the purifications compatible with this QBER.
\item We also need to consider all the measurements that can be done on Eve's part of the purification.
\end{itemize}
\subsection{Computation of the purification $\ket{\Psi_{ABE}}$}
In the entangled based scheme of BB84, Alice prepares an EPR state $\rho^0_{AB}=\ket{\Phi^+}\bra{\Phi^+}$ (where $\ket{\Phi^+}=\frac{\ket{00}+\ket{11}}{\sqrt{2}}$ is a Bell state) and sends one half to Bob. Due to Eve's action during the transmission, Alice and Bob now hold a noisy version of $\rho^0_{AB}$ that we denote by $\rho_{AB}=\mathcal{E}(\rho^0_{AB})$. Following \cite{renner_gis}, the security of BB84 can be studied without loss of generality on attacks for which the state $\rho_{AB}$ is Bell diagonal. We can write
\begin{eqnarray}\label{rhoab}
\rho_{AB}&= & \alpha \kb{\Phi^+}+\beta \kb{\Phi^-}+\notag\\
&&\gamma \kb{\Psi^+}+ \delta \kb{\Psi^-}\\
&& \mbox{with\ \ \ }\alpha+\beta+\gamma+\delta=1, \notag
\end{eqnarray}
where $\ket{\Phi^\pm}=\frac{\ket{00}\pm\ket{11}}{\sqrt{2}}$ and $\ket{\Psi^\pm}=\frac{\ket{01}\pm\ket{10}}{\sqrt{2}}$.\\
During the protocol, Alice and Bob use a small fraction of the raw key to estimate the QBER of their channel. Let $Q_0$ and $Q_1$ be the QBER measured by Alice and Bob when they measure in $\{\ket{0},\ket{1}\}$  and $\{\ket{+},\ket{-}\}$ respectively. We can write the relation between $Q_0$, $Q_1$ and the eigenvalues of $\rho_{AB}$ as
\begin{eqnarray}\label{qber}
Q_0&=&\bra{01}\rho_{AB}\ket{01}+\bra{10}\rho_{AB}\ket{10}=\gamma+\delta,\\
Q_1&=&\bra{+-}\rho_{AB}\ket{+-}+\bra{-+}\rho_{AB}\ket{-+}=\beta+\delta. \notag
\end{eqnarray}
If Alice and Bob measure a different value for $Q_0$ and $Q_1$, it gives them a clue that the channel has been tampered with. We therefore consider that $Q_0=Q_1=Q$. If we keep $\alpha\in[1-2Q,1-Q]$ (for $Q\in[0,1/2]$) as a free parameter, the state shared by Alice and Bob depends only on $\alpha$ and $Q$:
\begin{eqnarray}
\rho_{AB}&= & \alpha \kb{\Phi^+}+(1-Q-\alpha) \kb{\Phi^-}+\\ && (1-Q-\alpha) \kb{\Psi^+}+(2Q-1+\alpha) \kb{\Psi^-}.\notag
\end{eqnarray}
Eve has access to a purification $\ket{\Psi_{ABE}}$ of the state $\rho_{AB}$. The Schmidt purification can be obtained very easily from the orthonormal decomposition of $\rho_{AB}$:
\begin{eqnarray}
\ket{\Psi_{ABE}}&=& \sqrt{\alpha} \ket{\Phi^+}_{AB}\ket{\Phi^+}_{E}+\sqrt{1-Q-\alpha} \ket{\Phi^-}_{AB}\ket{\Phi^-}_{E}\notag\\
&&+\sqrt{1-Q-\alpha} \ket{\Psi^+}_{AB}\ket{\Psi^+}_{E}\\
&&+\sqrt{2Q-1+\alpha} \ket{\Psi^-}_{AB}\ket{\Psi^-}_{E}.\notag
\end{eqnarray}
A purification is not unique but we can choose this one without loss of generality since any purification of $\rho_{AB}$ can be obtained from $\ket{\Psi_{ABE}}$ with a suitable unitary acting on Eve's part of the purification. This unitary can be appended to the measurement performed by Eve after the interactions so we can safely ignore it.

\subsection{Optimization of $I_{AE}$}
To optimize her information, Eve wants to use the fact that the basis information will be revealed during the post-processing phase of the protocol. Even though she doesn't have access to a quantum memory and thus can not wait until she has this information to perform a measurement on her ancilla, she can choose her POVM in such a way that the post-measurement information will increase her knowledge on the raw key. We use a method similar to the one used in \cite{state_dis} in the case of state discrimination where it was argued that the most general measurement strategy for Eve was to use a POVM $\{M_{x_0x_1}\}_{x_0x_1=00,01,10,11 }$ with four possible outcomes $x_0x_1$. When Eve gets the measurement result $x_0x_1$, she waits for the basis information $\theta$ to be revealed so that she can choose $x_\theta$ as her guess.\\

The probability that Eve measures a certain value $k$ when Alice has obtained the result $x$ in the basis $\theta$ can be written as
\begin{eqnarray}
p(K=k|X=x,\Theta=\theta)&=& Tr(M_k\rho_E^{x\theta}),
\end{eqnarray}
where $\rho_E^{x\theta}$ represents Eve's part of the purification when Alice has obtained the result $x$ after a measurement in the basis $\theta$. We can compute this state from $\rho_{ABE}=\kb{\Psi_{ABE}}$:
\begin{eqnarray}
\rho_E^{x\theta}=\frac{\mbox{Tr}_{AB}[H^\theta\kb{x}{H^\theta}\otimes\mathbb{I}_B\otimes\mathbb{I}_E.\rho_{ABE}]}{\mbox{Tr}[H^\theta\kb{x}{H^\theta}\otimes\mathbb{I}_B\otimes\mathbb{I}_E.\rho_{ABE}]}.
\end{eqnarray}
The optimization problem is now reduced to the computation of the optimal POVM that maximizes the mutual information $I_{AE}$. It can be stated as:
\begin{eqnarray}
\mbox{maximize} & \mbox{\ \ \ \ \ \ } & I(X:K\Theta)\notag\\
\mbox{such that} & & \sum_i M_i = \mathbb{I}_4\\
&& \forall i \mbox{\ ,\ \ } M_i\geqslant 0 \notag
\end{eqnarray}
We solved this SDP problem numerically with the help of  \texttt{CVX} \cite{cvx,cvx2} and {\sl SDPT3} \cite{sdp} in MATLAB. In the next section we present the results we obtained when we applied this method to three different QKD protocols: the BB84, six-state and SARG04 QKD protocols.

\section{Optimal memoryless attacks for different QKD protocols}\label{protocols}
\subsection{Optimal memoryless attacks on BB84}

\begin{figure}[t]
\begin{center}
\includegraphics[width=\columnwidth]{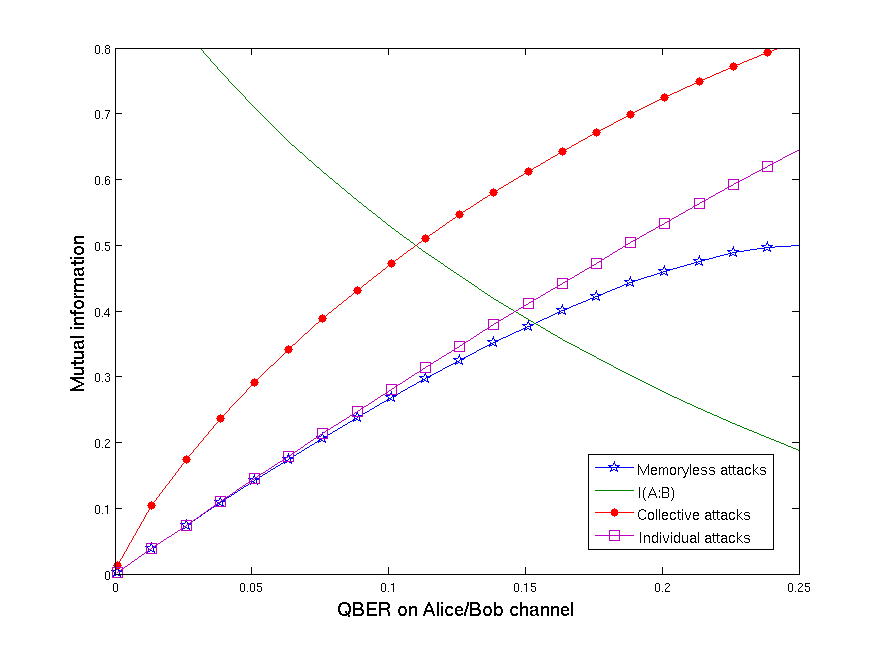}
\caption{Mutual information $I_{AB}$ and $I_{AE}$ in the BB84 protocol against the QBER for several attack models}
\label{plot}
\end{center}
\end{figure}

When applied to the BB84 QKD protocol, our optimization gives us a numerical representation of the function $I_{AE}$ for all $Q\in[0,1/4]$. It turns out that this result corresponds exactly to the mutual information between Alice and Eve that was computed in \cite{lutkenhaus_97} where Eve was allowed to perform a general POVM measurement directly on the flying qubit. The two methods agree on the optimal memoryless attack and this gives us the expression of the mutual information between Alice and Eve as it was computed in \cite{lutkenhaus_97}:
\begin{eqnarray}
I_{AE}&=&\frac{1}{2}+\frac{\Lambda[1+\epsilon(Q)]-\Lambda[\epsilon(Q)]}{2(1+\epsilon(Q))}\\
\mbox{with\ \ \ }\epsilon(Q)&=&\left(\frac{1-\sqrt{8Q(1-2Q)}}{1-4Q}\right)^2
\end{eqnarray}

In FIG. \ref{plot}, we have plotted the mutual information $I_{AB}$ against $I_{AE}$ for three different attack models: individual attacks, collective attacks and the optimal memoryless attacks on BB84.

We optimize Eve's strategy on all the accessible parameters: the choice of the purification and the measurement setting. Since we consider all the possible purifications and use the most general strategy for the measurement,  the result of our optimization is the optimal attack without a quantum memory.

The memoryless attack is always less effective than individual attacks and can never provide Eve with full information on the raw key: indeed, the mutual information $I_{AE}$ reaches its maximum of $1/2$ for $Q=0.25$. However we can see that the individual attacks (which require a quantum memory) do not provide Eve with much more information than the optimal memoryless attack.

It is well known that BB84 is secure against collective attacks up to a QBER $Q\approx11\%$. If you restrict the eavesdropper to a memoryless attack, we find that the BB84 protocol is then secure up to a QBER of $15.4\%$, the same value that was computed in \cite{lutkenhaus_97}. This is less than one point more than the $14.6\%$ corresponding to the individual attacks.

\subsection{SARG04}

\begin{figure}[t]
\begin{center}
\includegraphics[width=\columnwidth]{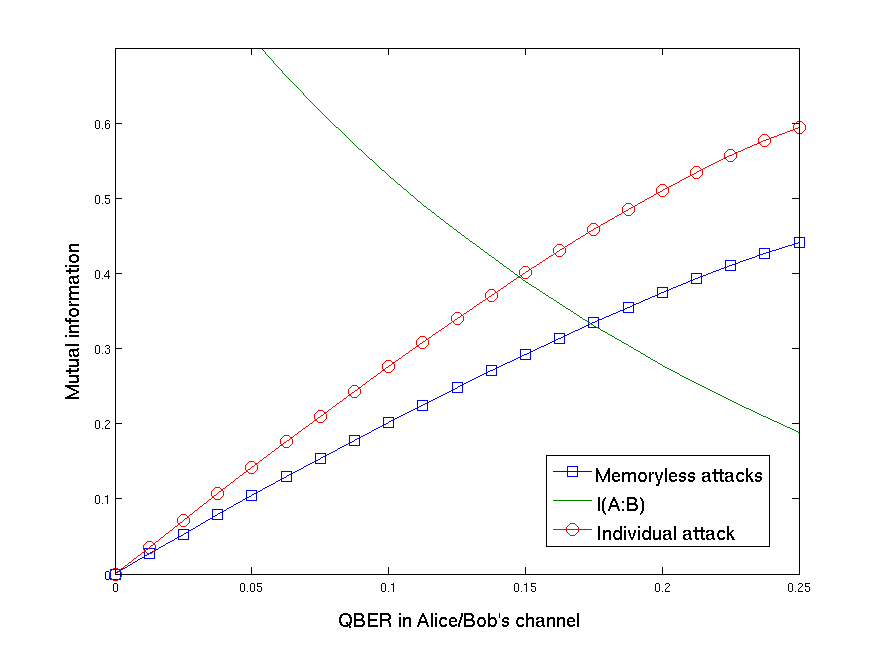}
\caption{Mutual information $I_{AB}$ and $I_{AE}$ in the SARG04 protocol against the QBER for several attack models}
\label{sarg}
\end{center}
\end{figure}

The SARG04 protocol \cite{SARG04} uses the same quantum states as BB84 but with a different encoding of information. In this protocol, Alice prepares a state $H^\theta\ket{x}$ where the classical bit is represented by $\theta$ instead of $x$ as in BB84. This means that the states $\ket{0}$ and $\ket{1}$ code for the classical bit $"0"$ and the states $\ket{+}$ and $\ket{-}$ code for the classical bit $"1"$. For example, if Alice chooses the classical bit $"0"$ and encodes it with the state $\ket{0}$, in the sifting phase she can announce $(\ket{0},\ket{+})$ to Bob. This does not give any information to Eve but it gives Bob full information about the classical bit if he measured in the basis $\ket{+},\ket{-}$ and got the result $"-"$.\\

From the point of view of Eve, the attack is the same that the one she does on the BB84 protocol. The difference with BB84 lies in the state $\rho_{AB}$. Indeed, if we use the same notations as for BB84 with $\rho_{AB}= \alpha \kb{\Phi^+}+\beta \kb{\Phi^-}+\gamma \kb{\Psi^+}+ \delta \kb{\Psi^-}$ we can follow \cite{branciard_05} and write:
\begin{eqnarray}
\alpha+\beta&=&1-Q\\
\gamma+\delta&=&Q\notag
\end{eqnarray}
so that we get:
\begin{eqnarray}
\rho_{AB}&= & \alpha \kb{\Phi^+}+(1-Q-\alpha) \kb{\Phi^-}+\\ 
&& (1-\frac{3Q}{2}-\alpha) \kb{\Psi^+}+(\frac{5Q}{2}-1+\alpha) \kb{\Psi^-}.\notag
\end{eqnarray}
After an optimization of the mutual information $I_{AE}$ over all the POVMs that Eve can use to measure her ancilla, we have computed that a key can be extracted for a QBER of less than $17.5\%$ against a memoryless adversary compared to $14.8\%$ for individual attacks \cite{branciard_05}. The mutual information between Alice and Eve for the individual attacks and the memoryless attacks are plotted on FIG. \ref{sarg}.

\subsection{Optimal memoryless attacks on the six-state protocol}

\begin{figure}[t]
\begin{center}
\includegraphics[width=\columnwidth]{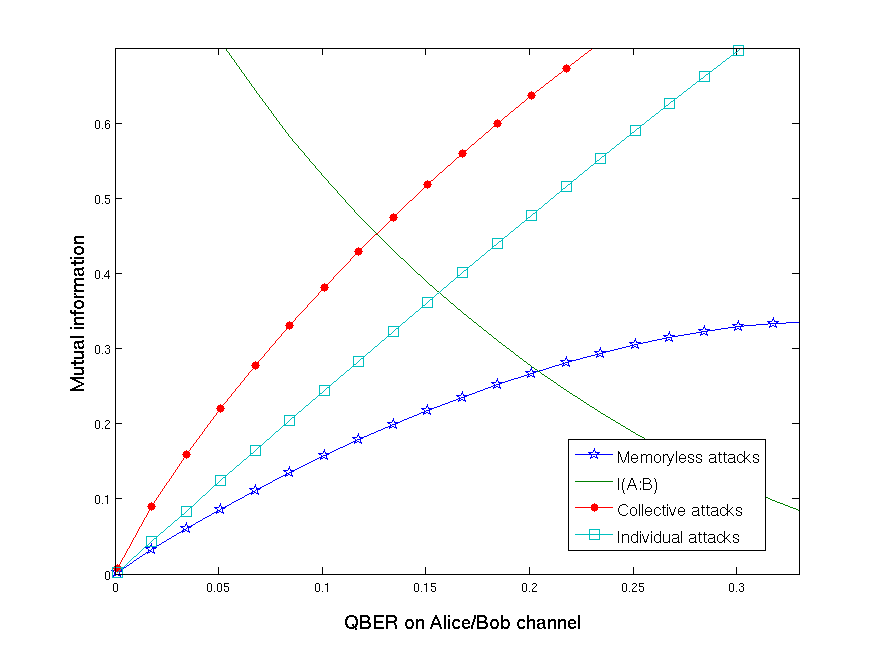}
\caption{Mutual information $I_{AB}$ and $I_{AE}$ in the six-state protocol against the QBER for several attack models}
\label{plot_sixstate}
\end{center}
\end{figure}

The six-state protocol \cite{lo_01,bechmann_99} is an extension of the BB84 protocol where three bases are used instead of only two. In this case, Alice can choose between the basis $\{\ket{0},\ket{1}\}$, $\{\frac{\ket{0}+\ket{1}}{\sqrt{2}},\frac{\ket{0}-\ket{1}}{\sqrt{2}}\}$ or $\{\frac{\ket{0}+i\ket{1}}{\sqrt{2}},\frac{\ket{0}-i\ket{1}}{\sqrt{2}}\}$ to encode her state. As a consequence the probability that Alice and Bob choose the same basis is only $1/3$ compared to $1/2$ in BB84: during the sifting phase $2/3$ on the bits have to be discarded. The advantage of this protocol is that its symmetry simplifies the analysis of its security and reduces the amount of information gained by Eve for a given QBER compared to BB84. Indeed, it was proven in \cite{lo_01} that the six-state protocol can produce a secret key up to a QBER of $12.6\%$ against the most general attacks.\\

The optimal memoryless attack on the six-state protocol follows the same procedure as the one described for BB84. Without loss of generality, we study the security of the six-state protocol against attacks for which the state $\rho_{AB}$ is diagonal in the Bell basis and can be written as equation (\ref{rhoab}). Since Alice and Bob can compute a QBER for three different bases, we get one additional relation between the diagonal coefficients and the QBER compared to equations (\ref{qber}). With the additional information that Alice and Bob measure the same QBER on each basis (the contrary would be a proof of tampering) we can write:
\begin{eqnarray}
\rho_{AB}&=&  (1-\frac{3Q}{2}) \kb{\Phi^+}+ \frac{Q}{2}\kb{\Phi^-}+\\
  &&\frac{Q}{2} \kb{\Psi^+}+ \frac{Q}{2} \kb{\Psi^-}.\notag
\end{eqnarray}

From this expression it is easy to get a purification $\ket{\psi_{ABE}}$:
\begin{eqnarray}
\ket{\psi_{ABE}}&=& \sqrt{1-\frac{3Q}{2}} \ket{\Phi^+}_{AB}\ket{\Phi^+}_{E}+\sqrt{\frac{Q}{2}} \ket{\Phi^-}_{AB}\ket{\Phi^-}_{E}\notag\\
&&+\sqrt{\frac{Q}{2}} \ket{\Psi^+}_{AB}\ket{\Psi^+}_{E}+\sqrt{\frac{Q}{2}} \ket{\Psi^-}_{AB}\ket{\Psi^-}_{E}.\notag
\end{eqnarray}
It is then possible to optimize the mutual information $I_{AE}$ on all the POVMs that Eve can use to measure her system. The result of this optimization is plotted on FIG. \ref{plot_sixstate} along with the mutual information for the collective \cite{lo_01} and individual attacks \cite{bechmann_99} of the six-state protocol. We find that a secret key can be extracted for a QBER of less than $20.4\%$ against a memoryless adversary compared to $12.6\%$ and $15.6\%$ for collective and individual attacks respectively.

\section{Conclusions}
We have shown how to construct the optimal memoryless attacks on BB84, six-states and SARG04 with an optimization of both the interaction U and the POVM used by Eve. Our result confirms the optimality of the previous bound of $15.4\%$ derived in \cite{lutkenhaus_97} in the case of BB84. We also provide new bounds for the six-state and SARG04 protocols against a memoryless adversary: the QBER over which no key can be extracted is increased to $20.4\%$ and $17.6\%$ respectively.\\

In this realistic model of a memoryless adversary, our work provides a quantitative estimate of the trade-off between the desired confidence on the security of the protocol (unconditional security or memoryless security model) and the achievable secret key rate.\\

Furthermore, the situation where the eavesdropper doesn't have access to a quantum memory is an extreme case of a more general security model where the eavesdropper is allowed to use a noisy memory. In the future, it will be interesting to study how the security bounds of QKD protocols evolve with the amount of noise in the eavesdropper's quantum memory. This model could also be used to prove the security of other protocols like the differential phase shift \cite{inoue_02} or continuous variables \cite{grosshans_03} protocols  against a memoryless adversary.\\

AB and RA acknowledge support from the French National Research Agency project FREQUENCY (ANR-09-BLAN-0410-01) and from the European Union through the project QCERT (FP7-PEOPLE-2009-IAPP). AL acknowledges support from the European Union through the  ERC Starting Grant PERCENT.

\bibliography{bibliography.bib}

\end{document}